\documentclass[twocolumn]{elsart3p}

\usepackage{epsfig}

\begin{document}

\begin{frontmatter}



\title{The characterization of a hybrid telescope detector for low energy alpha detection}


\author[label1,label2]{Rajkumar Santra},
\author[label1]{Subinit Roy},

\address[label1]{Saha Institute of Nuclear Physics, 1/AF Bidhan Nagar, Kolkata-700064, India}
\address[label2]{Homi Bhabha National Institute, Anushaktinagar, Mumbai-400094, India}

\begin{abstract}
Characterization of a hybrid ${\it telescope}$ with gas transmission detector ($\Delta$E) and a solid state stop detector (E) has been fabricated for detection of low energy  $\alpha$ particles between 5 to 1 MeV.   
The detector is developed for utilization in the study of alpha excitation function in (p.$\alpha$) reaction.   
The gas ionization chamber, operated in axial fild mode, measures the differential energy loss ($\Delta$E), while the residual energies are measured by silicon detector. Particle
identification is realized by implementing $\Delta$E-E technique. The optimum sensitivity of the detector as a telescope has been studied down to a lowest energy value of
0.89 MeV $\alpha$-particles with suitable combination of electric field and pressure or E/p value in the ionization region.

\end{abstract}

\end{frontmatter}

\section{Introduction}

The term ${\it telescope}$ in experimental nuclear physics stands for a combination of two charged particle detectors, capable of identifying the charge and mass (at least for light ions) of the incident particles. 
The combination works on the principle of partial energy loss in a thin detector ($\Delta$E) followed by a detector thick enough to stop the incident particle with residual energy (Stop E). Normally, 
the telescopes consist of a thin solid state detector as $\Delta$E and another thick silicon semiconductor detector as stopping detector. 
Normally, minimum 10 $\mu$m thick silicon detectors are used as a $\Delta$E detector in nuclear physics experiments. 
These thin silicon detectors have very poor energy resolution coming from large capacitance (C = $\frac{\epsilon _0 A}{d}$) and also have very short operational life as they are very prone to radiation damage. Also the 
$\Delta$E-E particle identification technique with silicon-silicon combination has a threshold dependence, governed by the thickness of the  $\Delta$E detector. A telescope with a 10 $\mu$m thick 
$\Delta$E detector is capable of particle identification for minimum 3-4 MeV $\alpha$-particles with low noise pulse processing electronics. 
For such low energy light ions, particle identification using a gas $\Delta$E detector is an useful alternative. Such a combination of gas $\Delta$E and solid state stop E detector 
for particle identification is termed as {\it hybrid} telescope setup. 
Gas $\Delta$E has an important advantage that its active thickness can be varied by simply adjusting the gas pressure and thus making it a transmission
type for low energy ions according to one's requirement.

In the present work, we designed,  fabricated and characterized a hybrid telescope for the detection of alpha particles with energy ranging from 5 MeV to about 1 MeV. We used standard radioactive $\alpha$-source for the testing and characterization.

\begin{figure}
\begin{center}
\includegraphics[height=8 cm, width=8.0 cm]{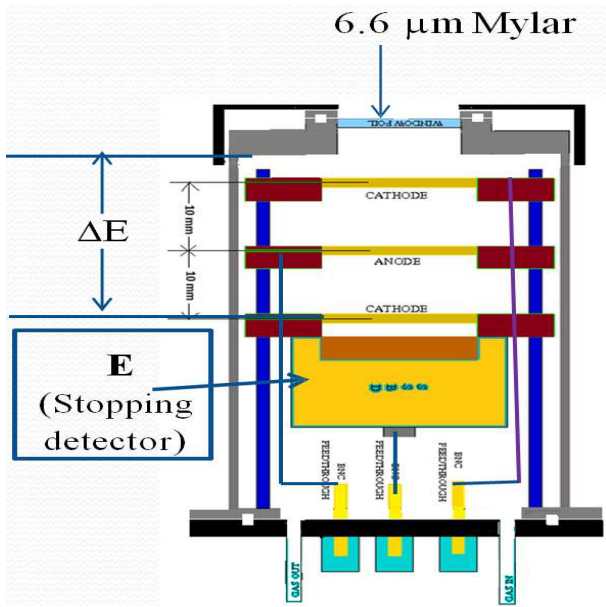}
\caption[The block diagram of E-$\Delta$E telescope.]{\label{hybrid} The block diagram of E-$\Delta$E telescope.}
\end{center}
\end{figure}

\section{Detector structure}
\begin{figure}
\begin{center}
\includegraphics[height=8 cm, width=8.0 cm]{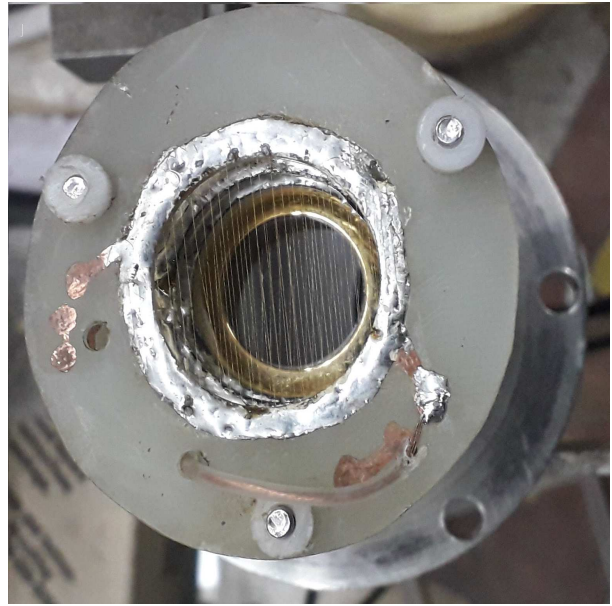}
\caption[The assembled wire frame.]{\label{top_view} The assembled wire frame.}
\end{center}
\end{figure}

In Fig.1, a schematic diagram of the hybrid telescope detector developed has been shown. The present hybrid telescope consists of an ionization chamber, followed by a Silicon surface barrier
detector of thickness 100 $\mu$m and 200 mm$^2$ active area from ORTEC. The gas $\Delta$E detector operates as an axial field ionization chamber. It has a three electrode geometry \cite{Jhingan} 
with a central anode sandwiched between two cathodes. As shown in Fig. ~\ref{hybrid}, the electrodes are mounted with their plane perpendicular to the direction of the
incoming particle. In an axial geometry, a uniform electric field covering the entire ionization volume can be maintained. The electrodes are fabricated using 25 $\mu$m
diameter gold plated tungsten wires. The stretched wires are soldered on a 1.6 mm thick annular printed circuit board (PCB) with a pitch of 1mm. Inter electrode distance 
is 10 mm giving a two-stage active length of 20 mm inside the gas volume. The wire frames are a cathode, a central anode frame, and another cathode wire frame. 
The distance between adjacent wire frames is 10 mm. Diameter of the active region is 20 mm. Fig. ~\ref{top_view} shows an assembled wire frame. The detector is operated within 
a range of electric field to pressure (E/p) ratio values of 1 to 8 Volt.cm$^{-1}$.mbar$^{-1}$ maintaining the operation in the ionization region. Typical working pressures 
of the detector are in the range of 24-288 mbar Isobutene(C$_4$H$_{10}$) gas, depending on the measurements to be performed. The particles enter the
detector through a Mylar window. Thickness of the Mylar window used is 6.6 $\mu$m. The signals from the anode of the gas detector and from the Silicon surface barrier 
detector are processed using the MSI-8 module of charge sensitive pre-amplifier and shaper amplifier combination from Mesytec. 

\section{Offline performance of the detector}
\begin{figure}
\begin{center}
\includegraphics[height=6 cm, width=8.0 cm]{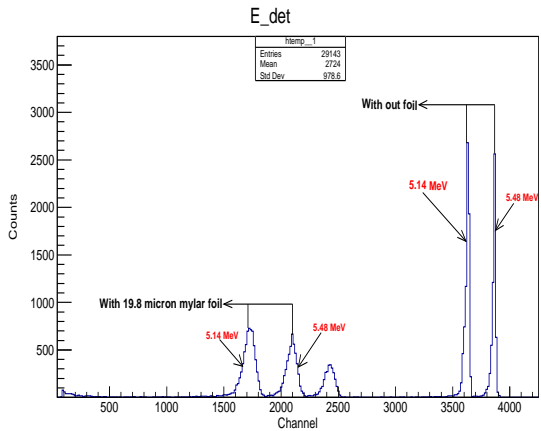}
\caption[The spectrum of three line $\alpha$-source without and with 19.8 Mylar foils.] {\label{spectrum} The spectrum of three line $\alpha$-source without and with 19.8 Mylar foils.}
\end{center}
\end{figure}

\begin{figure}
\begin{center}
\includegraphics[height=8 cm, width=8.0 cm]{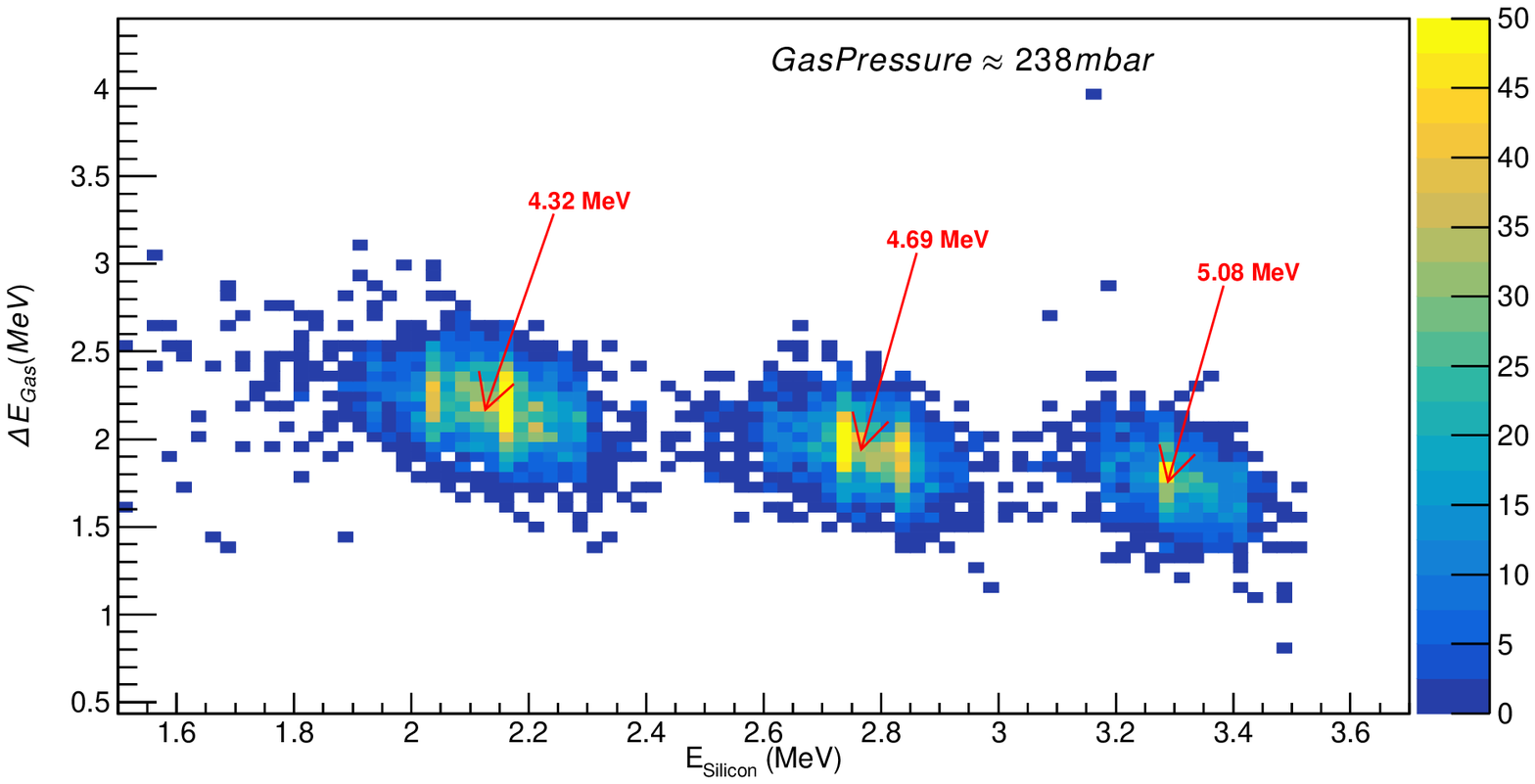}
\caption[The E(silicon)-$\Delta$E(gas) scattered plot for 238 mbar gas pressure] {\label{gas_si_one_foil} The two-dimensional scattered plot between $\Delta$E(gas) vs E(silicon) for 238 mbar gas pressure.}
\end{center}
\end{figure}

\begin{figure}
\begin{center}
\includegraphics[height=8 cm, width=8.0 cm]{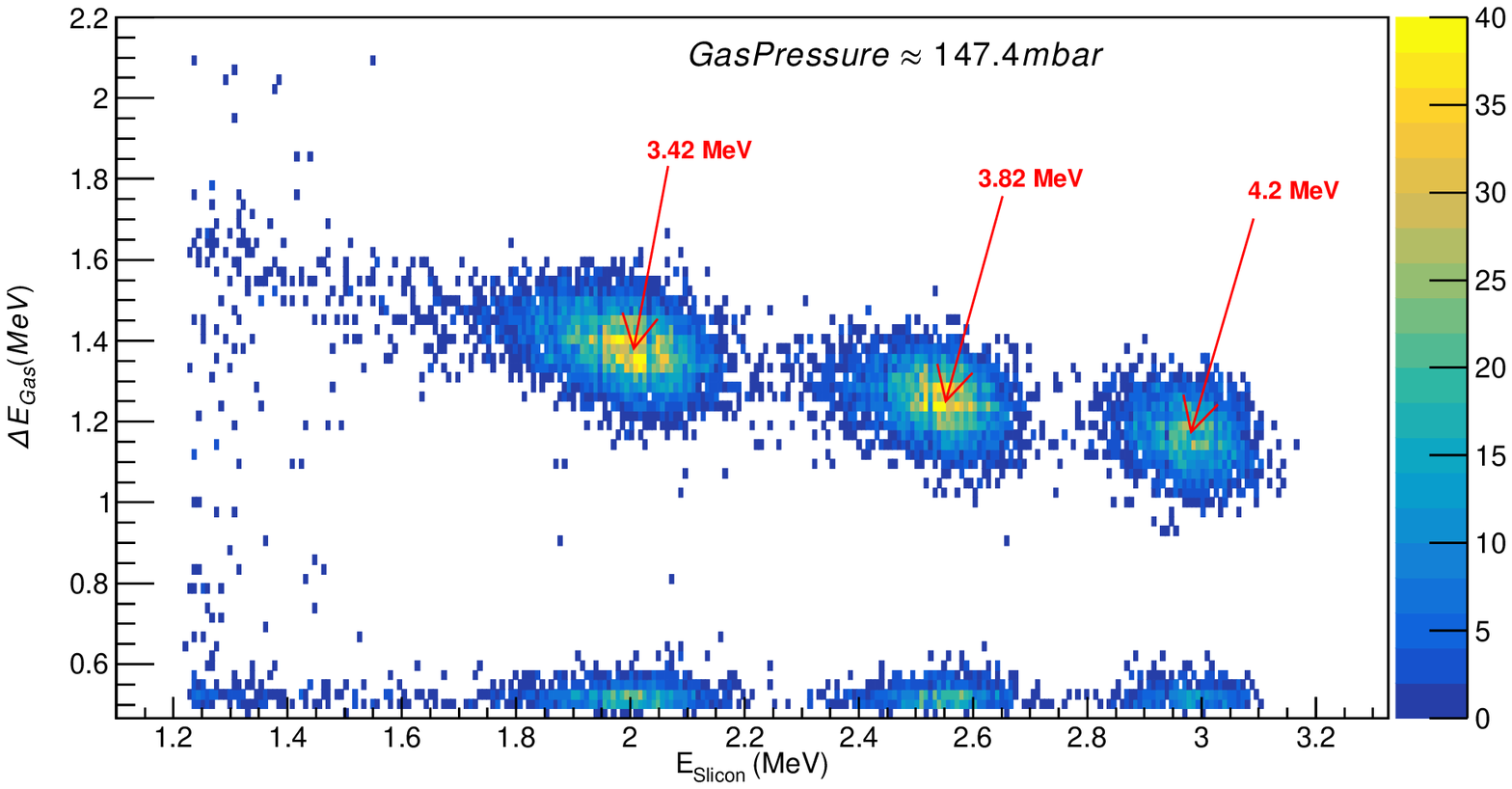}
\caption[The E(silicon)-$\Delta$E(gas) scattered plot for 147 mbar gas pressure]{\label{gas_si_two_foil} The two-dimensional scattered plot between $\Delta$E(gas) vs E(silicon) for 147.4 mbar gas pressure.}
\end{center}
\end{figure}

\begin{figure}
\begin{center}
\includegraphics[height=8 cm, width=8.0 cm]{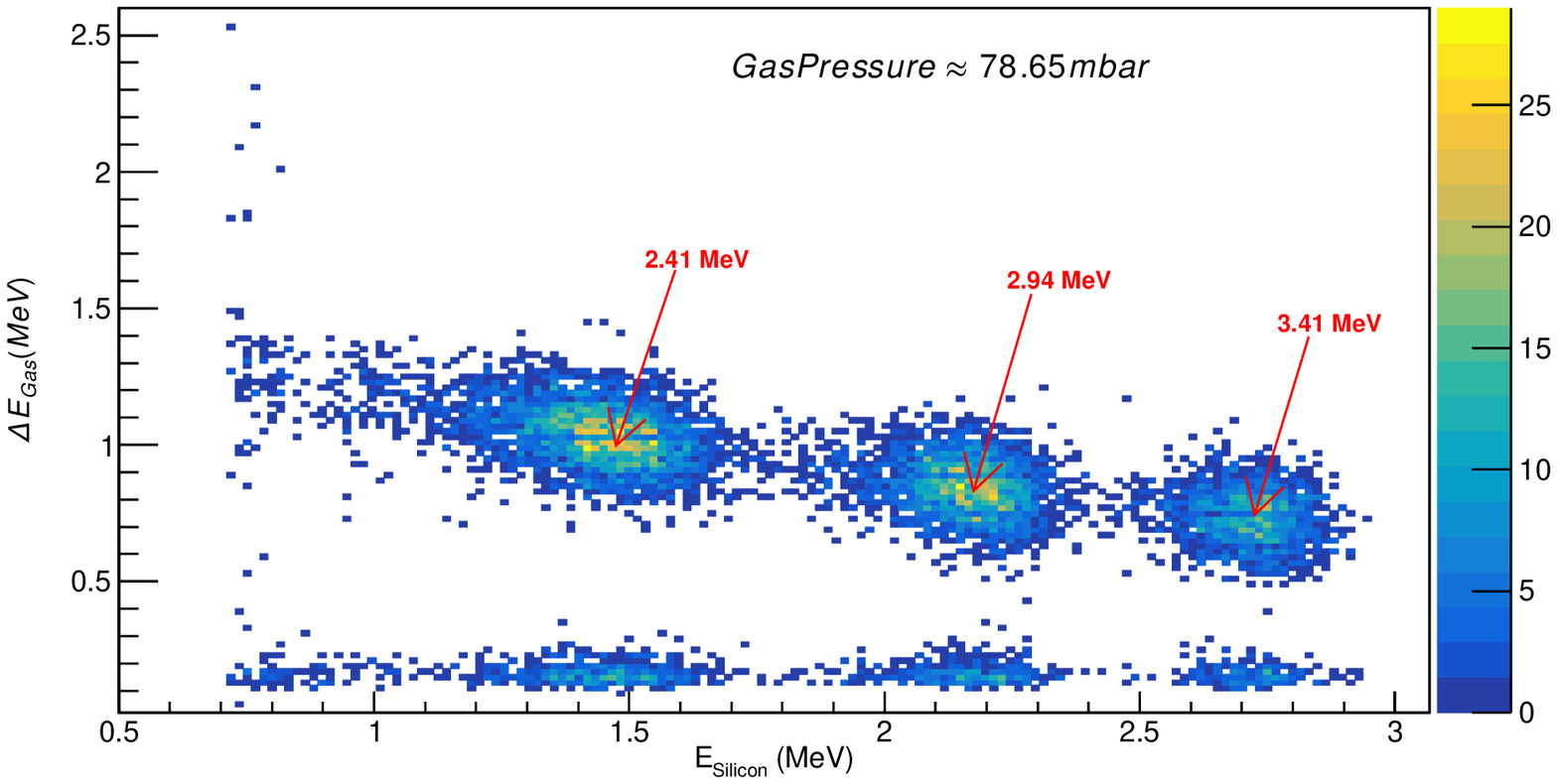}
\caption[The E(silicon)-$\Delta$E(gas) scattered plot for 78 mbar gas pressure]{\label{gas_si_three_foil} The two-dimensional scattered plot between $\Delta$E(gas) vs E(silicon) for 78.6 mbar gas pressure.}
\end{center}
\end{figure}

\begin{figure}
\begin{center}
\includegraphics[height=8 cm, width=8.0 cm]{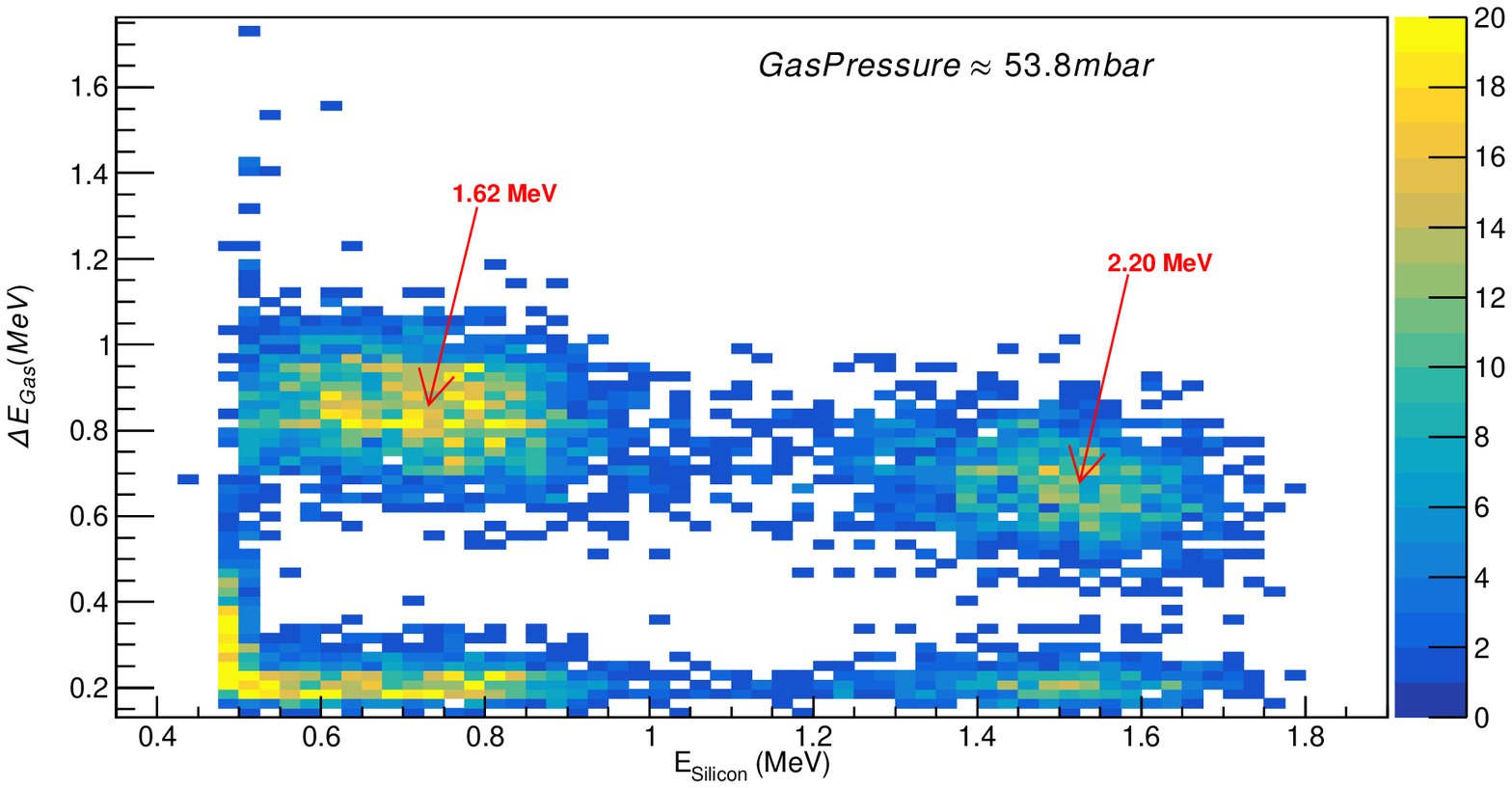}
\caption[The E(silicon)-$\Delta$E(gas) scattered plot for 53 mbar gas pressure]{\label{gas_si_four_foil} The two-dimensional scattered plot between $\Delta$E(gas) vs E(silicon) for 53.8 mbar gas pressure.}
\end{center}
\end{figure}

\begin{figure}
\begin{center}
\includegraphics[height=8 cm, width=8.0 cm]{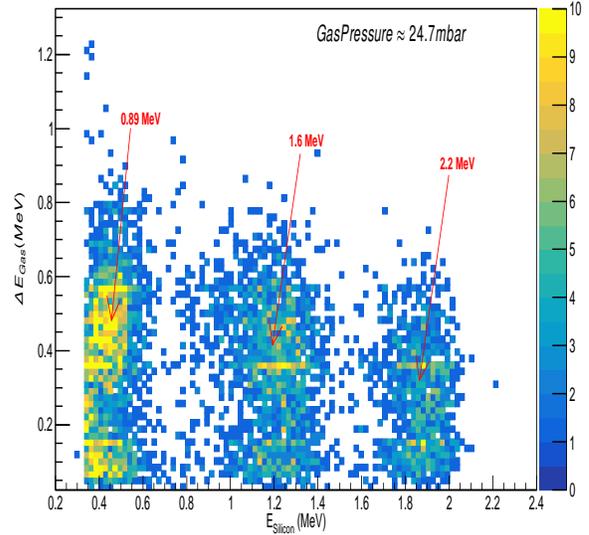}
\caption[The E(silicon)-$\Delta$E(gas) scattered plot for 24 mbar gas pressure]{\label{gas_si_four_foil_185} The two-dimensional scattered plot between $\Delta$E(gas) vs E(silicon) for 24.7 mbar gas pressure.}
\end{center}
\end{figure}

The detector is tested in the laboratory with a 3-line $\alpha$ source with average energy of 5.147 MeV from $^{239}$Pu, 5.485 MeV from $^{241}$Am and 5.79 MeV from $^{244}$Cm. Main motivation of the work is 
to study the performance of the hybrid detector as a telescope down to a low energy.Emphasis is on the lowest possible energy value for which the hybrid detector is sensitive as a telescope. 
To obtain the lower energy alphas, absorber foils have been used before the detector. 
Representative spectra without and with 19.8 $\mu$m absorber foil shown in Fig. \ref{spectrum} to get an idea of straggling effect of absorber. 
The calibration of the Silicon stop detector is done without introducing any gas in the detector and without any absorber foil. Resolution of the detector is $\sim$ 13.5 keV around 5 MeV energy. 
Also at each  stage of increasing thickness, the energy of the alpha particle incident on the detector has been determined from the stop detector without any gas in the active volume.
For different set of initial $\alpha$-energy values, the gas pressure has been optimized to obtain sufficient signal strength above the noise level from both the gas $\Delta$E and stop E detectors. 
As mentioned earlier, the primary motivation here is to achieve a reasonable sensitivity with this detector configuration in energy identification at very low energy. 
Thus, the optimum sensitivity of the detector as a telescope has been studied down to a lowest energy value of 0.89 MeV $\alpha$ particles with suitable combination of 
electric field and pressure or $E/p$ value in the ionization region. The resultant two- dimensional $\Delta$E vs. E spectra for different gas pressures in the 
$\Delta$E detector are shown in Figs.~\ref{gas_si_one_foil} ~\ref{gas_si_two_foil} ~\ref{gas_si_three_foil} ~\ref{gas_si_four_foil}. Even for an $\alpha$ energy 
of less than 1 MeV the response of the detector is reasonable as shown in Fig.~\ref{gas_si_four_foil_185}.

\begin{table}
\begin{center}
\caption[The gas pressures and electric fild.]{The gas pressures and electric fild between anode-cathode.}
\label{tab7}
\begin{tabular}{cccccccc} \hline \hline
Gas& pressure(P) & electric fild (E) &$\frac{E}{P}$& \\ 
&mbar&V. cm$^{-1}$&V. cm$^{-1}$. mbar$^{-1}$ \\ \hline
   & 238 & 270&1.13& &  \\
   &147.4  & 266&1.8 \\
Isobutene(C$_4$H$_{10}$) &78.6   & 200 &2.56& &  \\
    &53.6   & 200 &3.77& &  \\
    &24.7   & 200 &8.0& &  \\ \\\hline
\end{tabular}
\end{center}
\end{table}

\subsection{Conclusion}
In the present configuration and operating condition of the detector, we could detect $\alpha$-particles of energy just below 1 MeV.It is obvious from the figures that with decreasing incident alpha energy 
there is significant spread in energy due to the absorbers but still the three different energy bands from the source can be clearly identified.
 We used the detector in the laboratory with only $\alpha$-particles. 
Therefore the natural extension is to use it with the beam and check the particle identification capability of the detector. Also, by using a thinner window ($\sim$ 1-2 $\mu$m) 
and maintaining a stable gas flow mode in the detector the sensitivity of the detector can be further improved. 
Also in the flow mode with thin window of thickness about 0.5 $\mu$m, the detection of light heavy ions like carbon and oxygen can be achieved. 
Such a hybrid telescope detector is being developed for the use in the study of light ion fusion reactions at low energy.

\end{document}